\documentclass[sigconf]{acmart}
\AtBeginDocument{%
  }

\setcopyright{acmlicensed}
\copyrightyear{2025}
\acmYear{2025}
\acmDOI{XXXXXXX.XXXXXXX}
\acmConference[WSDM]{ The ACM International Conference on Web Search and Data Mining}{February 22–26, 2026}{Boise, Idaho, USA}
\acmISBN{978-1-4503-XXXX-X/2018/06}

\begin{document}

\title{SPARC: Soft Probabilistic Adaptive multi-interest Retrieval Model via Codebooks for recommender system}

\author{Jialiang Shi}
\authornotemark[1]
\affiliation{%
  \institution{Shanghai Dewu Information Group Co., Ltd.}
  \city{Shanghai}
  \country{China}}
  \email{shijialiang@dewu.com}

\author{Yaguang Dou}
\authornotemark[1]
\affiliation{%
  \institution{Shanghai Dewu Information Group Co., Ltd.}
  \city{Shanghai}
  \country{China}}
  \email{douyaguang@dewu.com}

\author{Tian Qi}
\authornotemark[1]
\affiliation{%
  \institution{Shanghai Dewu Information Group Co., Ltd.}
  \city{Shanghai}
  \country{China}}
  \email{qitian@dewu.com}

\renewcommand{\shortauthors}{Trovato et al.}

\begin{abstract}
Interest modeling in recommender system(RS) could improve user experience. Modeling multi-interests has arisen as a core problem in real-world RS. However, current multi-interest retrieval methods pose three major challenges: 1) Interests, typically extracted from predefined external knowledge, are invariant. Failed to dynamically evolve with users' real-time consumption preferences. 2) Online inference typically employs an over-exploited strategy, mainly matching users' existing interests, lacking proactive exploration and discovery of novel and long-tail interests. 3) Existed models fail to effectively mining multi-interests for users with few historical interactions. To address these challenges, we propose a novel retrieval framework named SPARC(Soft Probabilistic Adaptive Retrieval Model via Codebooks). Our contribution comes in two folds. First, The framework utilizes Residual Quantized Variational Autoencoder (RQ-VAE) to construct a learnable, discretized interest space. And to our best effort, it is the first time that achieves end-to-end joint training of the RQ-VAE with the industrial large scale recommendation model, mining behavior-aware interests that can perceive user feedback and evolve dynamically. Secondly, a probabilistic interest module that predicts the probability distribution over the entire dynamic and discrete interest space. This facilitates an efficient "soft-search" strategy during online inference, revolutionizing the retrieval paradigm from "passive matching" to "proactive exploration" and thereby effectively promoting interest discovery. Online A/B tests on an industrial platform with tens of millions daily active users, have achieved substantial gains in business metrics: +0.9\% increase in user view duration, +0.4\% increase in user page views (PV), and a +22.7\% improvement in PV500(new content reaching 500 PVs in 24 hours). Offline evaluations are conducted on open-source Amazon Product datasets. Metrics, such as Recall@K and Normalized Discounted Cumulative Gain@K(NDCG@K), also showed consistent improvement. Both online and offline experiments validate the efficacy and practical value of the proposed method.

\end{abstract}




\maketitle

\section{Introduction}
Real-world recommendation systems(RS) \cite{youtubeDNN} follow a multi-stage architecture, consisting of retrieval -> ranking -> re-ranking. The first stage, retrieval, focuses on locating candidate items with high recall and low response time. It decides the upper bound of recommendation quality, as subsequent stages primarily filter and rank the retrieved results. Recent research in RS have shown that user behavior modeling, which incorporates the user’s historical behavior sequences to learn implicit interests, results in significant performance gain \cite{mind, c-mifr, Atrank, DIN, DIEN}. 

Embedding-based retrieval \cite{emrfb, youtubeDNN, convws} has been proved to be one of the most effective and widely adopted model in industry, where users and candidate items are encoded as latent semantic embedding vectors, followed by inner-product function (or cosine function) as the satisfactions measurement. These architectures facilitate efficient retrieval via approximate nearest neighbor(ANN) \cite{ann1,ann2} search. These models are commonly referred to as "Two-Tower Models"(TT Model) \cite{ann2, youtubeDNN, samplebiascorrected}, where users and items are encoded separately in a user tower and an item tower, without interaction before the similarity computation. Despite effective and widely used, TT Models struggle to capture long-tail user interests due to the lack of early interaction between user and item features. Especially, dominant user interests tend to overwhelm the user embedding, resulting in fewer and inaccurate retrieved candidates from the torso and tail of a user’s interest distribution. This differs from the retrieval stage's primary motivation: maximizing the coverage of user's changeable multi-interests.

Pioneering works about multi-interests modeling in retrieval stage like MIND \cite{mind} and ComiRec \cite{c-mifr} have demonstrated the effectiveness of maximizing user interest coverage through dynamic routing and self-attention \cite{vaswani2017attention, Atrank}, setting a trend for research in this field. Despite significant progress, current multi-interest retrieval methods still face two fundamental challenges in industrial applications. First issue is the static and independent interest representations. Most existing solutions rely on a two-stage process, decoupled the definition of the interest space \cite{trinity} from the downstream recommendation task. Some approach directly uses human-defined category systems \cite{sim2020}, which are often inaccurate or lack appropriate granularity. Another approach first generates content features using external models (e.g., LLMs) and then constructs a static code-book using clustering or quantization techniques. This decoupled optimization leads to a "semantic-behavior gap": the static interest representations cannot be dynamically adjusted based on users' real-time consumption preferences, making them suboptimal for the final business objectives(e.g., click-through rate, user visit duration).
Second is the dilemma of passive matching versus exploration in the retrieval process. When deployed online, existing models often employ a greedy inference strategy. Specifically, the system typically performs a "hard-search" \cite{sim2020} using only a few of the user's dominant interests derived from their interaction history. This mechanism over-exploits the user's mainstream interests while neglecting their potential, emerging, or long-tail interests, thereby limiting the diversity \cite{ssd21} and novelty of the recommendations and harming the user's discovery experience. This pattern also exacerbates the long-standing cold-start problem in recommendation, where it is difficult to accurately infer the latent interests of new users or new content with sparse interaction histories.

\begin{figure}[htbp]
  \centering
  \includegraphics[width=\linewidth]{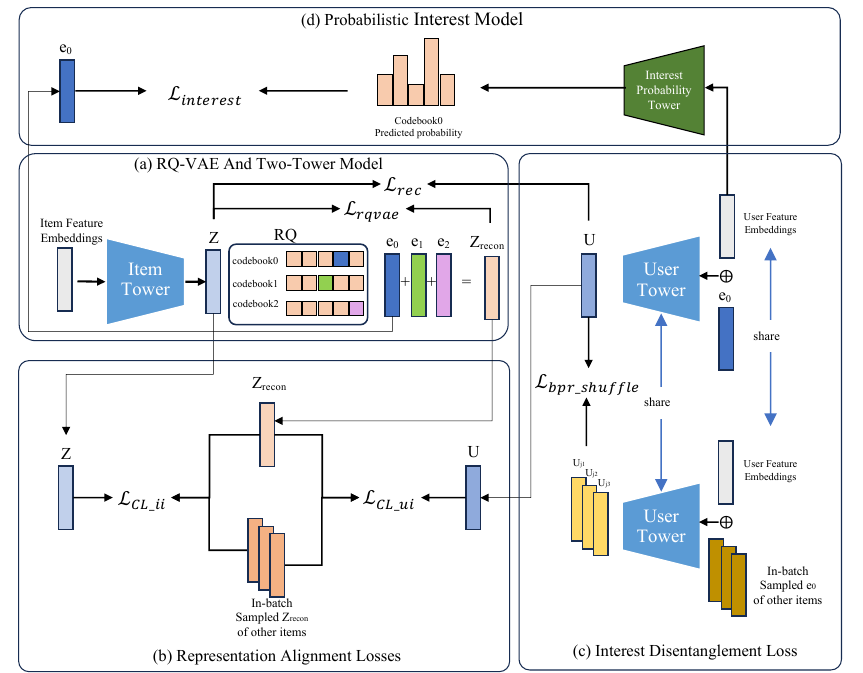}
  \caption{Overview of SPARC framework. It consists of four parts. a) RQ-VAE and Two-Tower Model. b) Representation alignment losses. c) Interest Disentanglement Loss. d) Probabilistic interest Model.}
  \label{retrievemodel}
\end{figure}

To systematically address the aforementioned challenges, this paper proposes SPARC(Soft Probabilistic Adaptive Retrieval Model via Codebooks for recomendation system), an end-to-end multi-interest retrieval framework, shown in Figure 1. The main contributions of this paper can be summarized in following two points.
\begin{itemize}
\item{\textbf{Joint training generative model with industrial retrieval model.}}
We redefine "interests" as dynamic, discrete codes learned within a semantic space through Residual Quantization (RQ) \cite{rqvae2022, rqinc2024}. We are the first to jointly train the RQ code-book with the main retrieval model in an end-to-end fashion. This allows interest codes to self-adjust according to the supervision signals from the recommendation task, thereby bridging the gap between representation learning and specific RS task. This paradigm transforms the content code-book from static cluster centers into "behavior-aware interest prototypes" that perceive user feedback and evolve dynamically, fundamentally solving the alignment problem between interest representation and downstream RS tasks.
To ensure that the end-to-end paradigm learns high-quality interest representations, we designed a sophisticated multi-task learning objective. It includes a BPR \cite{bpr2009} loss with random codeword shuffling and a dual contrastive learning loss \cite{simCLR}. These mechanisms work in synergy to learn interest code-book that are disentangled, robust, and aligned with their original semantics.
\item{\textbf{Probabilistic Interest exploration module.}}
Building on the high-quality discrete interest code-book space, we further designed a "soft-search" mechanism to innovate online serving. This mechanism, through a probabilistic interest module, predicts a user's interest distribution over all interests and executes parallel ANN retrievals. This effectively balances interest exploration and exploitation, upgrading the retrieval paradigm from "passive matching" to "proactive exploration", and significantly enhancing recommendation diversity and novelty.
\end{itemize}

We conducted comprehensive offline experiments and online A/B tests on a large-scale commercial recommendation system. The results show that SPARC not only excels in offline metrics, but also achieves significant relative improvements in key online business metrics (user visit duration +0.9\%, page views +0.4\%), demonstrating its effectiveness and practical value.

\section{Related Work}

\subsection{Multi-interest Recommendation}
The pioneering work MIND \cite{mind} adapt the dynamic routing mechanism from capsule networks \cite{capsuleNet} to cluster a user's historical behavior sequence into multiple vectors representing different interests. Building on this, ComiRec \cite{c-mifr} advanced this idea, typically incorporating a multi-interest extraction module and an aggregation module, and explored interest control mechanisms for online inference. Other works have explored different aggregation strategies, such as the self-attention mechanism \cite{vaswani2017attention}. These method could understand user's implicit multi-interests by representing a user with multiple representation vectors encoding the different aspects of the user's interests. However, these methods share common limitations. Both method incorporate sophisticated hand-craft network to learn the multi-interests semantic embedding. These embeddings are hard to verify accuracy from human understanding, resulting from the obscure relationship between user's real interests and learned interest embedding. Our SPARC framework is fundamentally different. We focus on learning the discrete structure of this semantic embedding space, incorporating RQ-VAE\cite{rqvae2022} model to bottleneck the semantic embedding space into discrete code-book space, which explicit cluster the semantic embedding interest space into countable and measurable interest sets, making generated code-book self-explanation. 

\subsection{Vector Quantization in Recommender Systems}
Vector Quantization (VQ) \cite{vq2011, vq2021} is a classic technique for data compression and representation. The basic idea is to approximate a high-dimensional vector using a codeword from a code-book. Building upon VQ, Product Quantization (PQ) \cite{pq2011} and Residual Quantization (RQ) \cite{rqvae2022, rqinc2024} were developed. Among them, RQ progressively approximates the original vector using multi-level quantization, where each level encodes the quantization residual from the previous one, thus achieving higher precision with smaller code-books \cite{rqinc2024}. In recent years, the application of VQ techniques in recommender systems has become increasingly prevalent. A recent survey \cite{liu2024vectorquantizationrecommendersystems} categorizes its applications into two types: (1) Efficiency-oriented applications, which leverage VQ for ANN retrieval acceleration and model compression to address scalability challenges in large-scale systems \cite{johnson2017billionscale}; and (2) Quality-oriented applications, which focus on improving recommendation performance, for example, by using VQ for feature enhancement, multi-modal alignment, or generating discrete item tokens for generative recommendation models \cite{journals/pami/JegouDS11}. Early work, particularly in generative recommendation, often used a pre-trained VQ-VAE or RQ-VAE to compress embedding space to code-book space, whose code-book was fixed before the recommendation model training. This two-stage paradigm decouples the tokenization process from the recommendation task itself, leading to a mismatch between the code-book's optimization objective (usually reconstruction loss) and the recommendation task's objective (e.g., CTR prediction). Consequently, the learned discrete codes fail to effectively capture the collaborative filtering signals crucial for recommendation. To address this, the paper proposes end-to-end joint training, a key innovation that enables gradients from the recommendation task to directly update the code-book. This ensures that the code-book's definition of "user interest" is explicitly optimized for recommendation performance, bridging the gap between tokenization and downstream recommendation tasks.

\subsection{End-to-End Recommendation Models}
Integrating traditional multi-stage recommendation pipelines (e.g., retrieval, ranking) into a unified, end-to-end trainable system is a significant trend in the current recommendation field \cite{naumov2019learning}. Among these, generative recommendation, which directly generates item IDs, is an emerging direction receiving much attention \cite{conf/sigir/00050LH0Z25}.
In this area, ETEGRec \cite{conf/sigir/00050LH0Z25} is an important state-of-the-art work. It also unifies item tokenization (using RQ-VAE) and the recommendation model (a T5-like generative model) into an end-to-end framework for joint optimization. The success of ETEGRec has demonstrated the great potential of jointly optimizing the tokenizer and the main recommendation model. Our work is highly consistent with ETEGRec in its core philosophy, but differs in application scenarios and technical implementation. ETEGRec focuses on the generative recommendation paradigm, whereas our proposed SPARC framework applies this core idea to the two-tower retrieval paradigm, which is more efficiency in industrial settings. This makes the idea more applicable to existing large-scale retrieval systems, giving it greater practical significance. Furthermore, ETEGRec couples the two modules through "sequence-item alignment" and "preference-semantic alignment," whereas our paper designs a different set of alignment loss functions (BPR loss and contrastive learning losses) for the retrieval task.

\subsection{Contrastive Learning in Recommender Systems}
Contrastive learning (CL)\cite{simCLR}, which pulls positive pairs closer while pushing negative pairs apart in representation space, has emerged as a powerful paradigm for learning high-quality representations. In recommender systems, CL has been applied in diverse settings: CLRec \cite{conf/cikm/ZhouWZZWZWW20}  leverages CL to mitigate exposure bias; AdaGCL \cite{conf/kdd/Jiang0H23} and ReHCL \cite{conf/aaai/WangZZWJ24} integrate contrastive objectives into graph neural networks to enhance node representations; and SGL \cite{conf/sigir/WuWF0CLX21}  constructs positive samples by generating augmented views (e.g., node dropping) on the user–item bipartite graph. Building on these advances, we design two contrastive losses, user-item (UI-CL) and item-item (II-CL), to ensure that item representations remain consistent with their original counterparts after RQ-VAE quantization and reconstruction, thus benefiting downstream recommendation. Additionally, We introduce a BPR loss with random codeword shuffling(RCS), tailored to our discrete interest space, which could enforce the separability of user vectors induced by different interest codewords.

\section{Method} 
 In this section, we will detail the SPARC framework, shown in Figure 1. First, we will introduce the overall architecture. Subsequently, we will delve into four core components: (1) the end-to-end residual quantization module, responsible for constructing the dynamic discrete interest space; (2) the probabilistic interest-aware user representation; (3) the advanced multi-task optimization framework that ensures the model learns high-quality interest representations; and (4) the probabilistic interest exploration mechanism that revolutionizes the online serving paradigm. 

 \subsection{Overall Architecture} 
 The proposed SPARC architecture is based on a two-tower model structure, as shown in the figure 1. The architecture primarily consists of an user tower, an item tower, a learnable residual quantization (RQ) code-book, and a probabilistic interest module. 

 \textbf{Item tower:} It takes raw item features (e.g., ID, category, multi-modal features) as input and encodes them into a high-dimensional dense item embedding vector $z$, through multi DNN layers. 

 \textbf{Residual Quantization Code-book and Reconstruction:} The item embedding $z$ is fed into a three-level RQ-VAE module, which each leval contains one code-book, each code-book consist of 256 codewords, dimension of each codewords is 64. This module quantize $z$ into a set of discrete code indices $$(\texttt{idx\_0}, \texttt{idx\_1}, \texttt{idx\_2})$$ and generates a reconstructed vector $z_{\text{recon}}$. Unlike traditional methods, this code-book is part of the model and is jointly trained end-to-end with the two-tower model. 

 \textbf{User Tower:} Takes the user static features (e.g., user age, user gender) and dynamic behavior sequence (e.g., recently clicked items) as input. Its key feature is the additional crossing and interaction: the reconstructed codeword vector $z_{\text{recon}}$ from the item side's RQ-VAE module interacts with the user's behavior sequence through target attention. It also undergoes deep crossing with other user static features and MLP hidden layers to generate a context-aware user embedding $u$ representing the user's current interest. 

 \textbf{Probabilistic Interest Module:} This module is independent of the main recommendation task and is responsible for predicting the user's interest probability distribution $P(c_1|u)$ over the first-level codebook. This distribution is used during the "soft-search" phase of online serving. 

 \textbf{Prediction and Optimization:} The model ultimately predicts the click-through rate (pCTR) by calculating the similarity between the user embedding $u$ and the item embedding $z$. It is optimized end-to-end using a multi-task loss function that includes a BPR loss with random codeword shuffling and dual contrastive learning losses. 

 \subsection{End-to-End Residual Quantization} 
 Residual Quantization is the core of this framework. It maps continuous item embedding vectors to a dynamic, discrete, and hierarchical code space and is trained end-to-end with the main model. We employ an RQ-VAE structure with $M=3$ levels of codebooks, with each level containing $K=256$ codewords of dimension 64. For a given item embedding $z$, the quantization process is as follows: 

 First-level quantization: $e_0=Q_0(z)$, where $Q_0$ finds the nearest codeword in the first-level codebook $\mathcal{C}_0$ to $z$. 

 The residual after the first level is $r_1=z-e_0$. 

 Second-level quantization: $e_1=Q_1(r_1)$, where $Q_1$ quantizes the residual $r_1$ in the second-level codebook $\mathcal{C}_1$. 

 The new residual is $r_2=r_1-e_1$. 

 Third-level quantization: $e_2=Q_2(r_2)$, performing the final quantization on the residual $r_2$. 

 Ultimately, the discrete representation of item $z$ is its set of indices in the three code-books, $({idx\_0}, {idx\_1}, {idx\_2})$, and its reconstructed vector is $z_{\text{recon}}=e_0+e_1+e_2$. The first-level codeword $c_{1, \texttt{idx\_1}}$ can be seen as the core interest of the item, while subsequent codewords capture finer-grained styles or attributes. 

 Joint optimization is a key aspect of this paper. The RQ-VAE's training loss, $\mathcal{L}_{\text{rqvae}}$, is integrated into the total recommendation model loss. $\mathcal{L}_{\text{rqvae}}$ itself comprises a reconstruction loss and a code-book commitment loss, formulated as: 

 \begin{equation} 
 \mathcal{L}_{\text{rqvae}} = \|z - z_{\text{recon}}\|_2^2 + \beta \sum_{k=0}^{2} \|sg(r_k) - e_k\|_2^2 
 \end{equation} 

 where $r_0=z$, $sg(\cdot)$ denotes the stop-gradient operation, and $\beta$ is a hyperparameter. Since $\mathcal{L}_{\text{rqvae}}$ is part of the total loss, gradients from the upper-level recommendation loss (e.g., $\mathcal{L}_{\text{BCE}}$) can backpropagate and update the codebooks $\mathcal{C}_0, \mathcal{C}_1, \mathcal{C}_2$. This forces the learning of the codebook vectors (i.e., the discretized "interests") to not only minimize reconstruction error but also to be beneficial for the final recommendation task, thereby achieving high alignment between representation learning and recommendation objectives. 

 \subsection{Probabilistic Interest-Aware User Representation} 
 The goal of the User Tower is to generate a dynamic user interest representation based on the user's historical behavior and the current candidate item. 

 \textbf{User Tower Structure:} The core of the User Tower is an attention module that uses the quantized information from the item side as a Query to perform a weighted aggregation of the user's behavior sequence. Specifically, the user's effective click sequence and its associated side-info are embedded and pooled to form a sequence representation $\mathcal{S}_{\text{user}} \in \mathbb{R}^{L \times D}$, where $L$ is the sequence length and $D$ is the dimension. The attention mechanism computes a context-aware user interest vector. 

 \textbf{Mixed Cross-Feature Strategy:} To enhance the model's robustness and generalization, a random mixing strategy is employed during training. For 50\% of the training samples, we use the first-level codebook vector $e_0$ as the query for the attention module; for the other 50\%, the full reconstructed vector $z_{\text{recon}}$ is used as the query. This design can be seen as a form of stochastic regularization: using only $e_0$ forces the model to learn to capture user preferences from coarse-grained interest signals, while using $z_{\text{recon}}$ allows the model to perform finer and more accurate matching. By training on signals of both granularities, the User Tower can learn representations that are both generalizable and precise. 

 \textbf{Architecture Modules:} In the feature fusion stage, the model adopts established industrial modules such as PEP-NET [25] and SE-Block [26] to achieve effective crossing and weighting of various user-side features (including attention output, user profile features, target item cross-features, etc.). 

 \subsection{Multi-Task Optimization Framework for Disentanglement and Alignment} 
 To effectively conduct end-to-end training and learn high-quality interest representations, we designed a joint optimization objective comprising multiple loss functions. The total loss function $\mathcal{L}_{\text{total}}$ is a weighted sum of the individual losses: 

 \begin{align}
 \mathcal{L}_{\text{total}} = &\mathcal{L}_{\text{BCE}} + \lambda_{\text{bpr}}\mathcal{L}_{\text{BPR\_shuffle}} + \lambda_{\text{interest}}\mathcal{L}_{\text{interest}} \nonumber \\ 
 &+ \lambda_{\text{ui}}\mathcal{L}_{\text{CL\_ui}} + \lambda_{\text{ii}}\mathcal{L}_{\text{CL\_ii}} + \lambda_{\text{rq}}\mathcal{L}_{\text{rqvae}} 
 \end{align} 

 Each loss is defined as follows: 

 \textbf{$\mathcal{L}_{\text{BCE}}$}: The main task loss, which is the standard binary cross-entropy loss for predicting the user's click behavior on the target item. 
 \begin{equation}
\mathcal{L}_{\text{BCE}} = -\mathbb{E}_{(u,i) \sim \mathcal{D}} \left[ y \log \sigma(\mathbf{u}^T \mathbf{z}) + (1 - y) \log (1 - \sigma(\mathbf{u}^T \mathbf{z})) \right]
\end{equation}
 where $\mathcal{D}$ is the training dataset, $y_{ui}$ is the ground truth label for user $u$ and item $i$, and $\sigma(\cdot)$ is the sigmoid function. 

 \textbf{$\mathcal{L}_{\text{BPR\_shuffle}}$ (Interest Disentanglement Loss):} 

This is a novel BPR loss designed to enhance the distinguishability of different interest codes. To enhance the separability of user vectors generated by different interest codewords, we designed a novel BPR loss with random codeword shuffling. For a positive pair $(u,i)$, we take the first-level codeword $c_{1, \texttt{idx\_1}}$ of item $i$ as its core interest representation. Within the same batch, we randomly sample $J$ core interest codewords $c'_{1,j}$ for $j=1, \dots, J$ from other items as negative samples. Then, we apply cross-attention between these codewords and user $u$'s historical sequence to generate one positive user vector $u_q$ and $J$ negative user vectors $u'_{q,j}$ for $j=1, \dots, J$. The BPR loss aims to maximize the margin between the scores of the positive and negative pairs: 
 \begin{equation} 
 \mathcal{L}_{\text{BPR\_shuffle}} = \sum_{(u,i) \in \mathcal{D}} \sum_{j=1}^{J} -\log \sigma(u_q^T z - (u'_{q,j})^T z) 
 \end{equation} 
 This loss function compels the User Tower to learn representations that are highly sensitive to the input interest codes, thereby enabling different interest codes to generate disentangled user vectors. In experiments, the weight $\lambda_{\text{bpr}}$ for this loss was set to 10, highlighting its importance. 

 \textbf{$\mathcal{L}_{\text{CL\_ui}}$ and $\mathcal{L}_{\text{CL\_ii}}$ (Representation Alignment Losses):} These are two in-batch contrastive learning losses. 

 $\mathcal{L}_{\text{CL\_ii}}$ aims to align the original item embedding $z$ with its reconstructed version $z_{\text{recon}}$, ensuring that the quantization process preserves the semantic information crucial for the recommendation task. 
 \begin{equation} 
 \mathcal{L}_{\text{CL\_ii}} = -\log \frac{\exp(\langle z, z_{\text{recon}} \rangle / \tau)}{\sum_{j \in \text{batch}} \exp(\langle z, z_{j, \text{recon}} \rangle / \tau)} 
 \end{equation} 

 $\mathcal{L}_{\text{CL\_ui}}$ ensures that the user vector $u$ aligns with the item's reconstructed vector $z_{\text{recon}}$, making the User Tower's output more robust to errors introduced by quantization. 
 \begin{equation} 
 \mathcal{L}_{\text{CL\_ui}} = -\log \frac{\exp(\langle u, z_{\text{recon}} \rangle / \tau)}{\sum_{j \in \text{batch}} \exp(\langle u, z_{j, \text{recon}} \rangle / \tau)} 
 \end{equation} 

 \textbf{$\mathcal{L}_{\text{interest}}$ (Interest Supervision Loss):} This loss is used to train an independent "interest probability tower." This module takes user features and behavior sequences as input and, through a self-attention network similar to ComiRec [6], outputs 256-dimensional logits. After a softmax layer, it yields the user's probability distribution over the 256 first-level interest codes, $P(\text{interest}|u)$. The training objective is to maximize the probability of the first-level code index $\texttt{idx\_0}$ corresponding to the positive item. 

 \subsection{Online Serving with Probabilistic Exploration} 
 During the online serving stage, SPARC employs an innovative "soft-search" strategy to effectively balance interest exploration and exploitation, directly addressing the greedy inference problem of traditional methods. The process is divided into two steps: 

 \textbf{Phase 1: Interest Prediction.} When a user request arrives, the pre-trained "interest probability tower" is first invoked to calculate the user's probability distribution $P(\text{interest}|u)$ over the 256 first-level interest codes. 

 \textbf{Phase 2: Parallel Retrieval.} 
 \begin{itemize}
    \item Based on the probability distribution, the Top-K most likely interest codes are selected (e.g., K=5 based on implementation). 
    \item For each of these K codes, the corresponding codeword vector $e_{0,k}$ is retrieved from the first-level codebook $\mathcal{C}_0$. 
    \item Each $e_{0,k}$, along with other user features, is separately fed into the User Tower to generate K different user embedding vectors, $u_1, u_2, \dots, u_K$, each representing a specific interest. 
    \item These $K$ user vectors are used to perform $K$ parallel ANN retrievals in the item corpus. 
    \item Finally, the results from the K retrieval paths are aggregated, deduplicated, and re-ranked to form the final recommendation list. The retrieval quota can be allocated based on the interest probabilities distribution, thus achieving a theoretically grounded and balanced exploration. 
\end{itemize}

 This mechanism enables the system to move beyond the user's single most dominant interest and actively explore multiple facets the user might be interested in, thereby improving the diversity and novelty of the recommendation results.

 \begin{table}[t]
\caption{Statistics of datasets.}
\small
\centering
\begin{tabular}{lc|clc|clc|cl}
\toprule
\multicolumn{1}{c}{Dataset} &
\multicolumn{1}{c}{\# users} &
\multicolumn{1}{c}{\# items} &
\multicolumn{1}{c}{\# interactions} \\
\midrule
\multicolumn{1}{c}{Amazon Books}  &
\multicolumn{1}{c}{459,133}  &
\multicolumn{1}{c}{313,966}  &
\multicolumn{1}{c}{8,898,041} \\

\multicolumn{1}{c}{Industrial Dataset}  &
\multicolumn{1}{c}{358,920,221}  &
\multicolumn{1}{c}{43,156,789}  &
\multicolumn{1}{c}{4,732,456,754} \\
\bottomrule
\end{tabular}
\label{tab:dataset_stats}
\end{table}

\section{Experiments}
\label{sec:experiments}

In this section, we conduct a series of offline experiments to comprehensively evaluate the effectiveness of our proposed SPARC framework. Our experiments are designed to answer the following key Research Questions (RQs):
\begin{itemize}
    \item \textbf{RQ1:} How does the overall performance of SPARC compare against state-of-the-art retrieval models?
    \item \textbf{RQ2:} As per its design, can SPARC effectively enhance recommendation novelty and the discovery of long-tail content? Can this explain its significant improvement on the new content consumption metric (PV500) in online experiments?
    \item \textbf{RQ3:} What are the respective contributions of SPARC's key components—specifically the dynamic RQ-VAE codebook and the probabilistic soft-search mechanism—to its final performance?
    \item \textbf{RQ4:} Does SPARC exhibit stronger generalization capabilities than other models for users with sparse interaction histories (the cold-start problem)?
\end{itemize}

\subsection{Experimental Setup}

\subsubsection{Datasets}
We use the 2023 version of the public \textbf{Amazon Product Data - Books}~\cite{amazon1, amazon2} for our experiments. This dataset contains rich user behaviors. Following the methodology of previous work~\cite{c-mifr}, we use the \textit{5-core} version, ensuring that each user and item has at least five interactions. We adopt the official \textit{Leave-Last-Out} splitting strategy to create training, validation, and test sets for each user's interaction sequence. For each interaction, we only consider those with a rating of 4.0 or higher as positive feedback. User history sequences are truncated to the 50 most recent actions. Also, we conduct experiment on our real-world datasets, which are about 4.7 billions instances. The statistics of the processed dataset are shown in Table~\ref{tab:dataset_stats}.


\subsubsection{Evaluation Protocol}
For each positive sample in the validation and test sets (i.e., the last interaction for each user), we associate it with 100 candidate items for evaluation. This set includes the one positive item and 99 negative items \textbf{randomly} sampled from the entire item pool that the user has never interacted with. To ensure reproducibility, the negative sampling process uses a fixed random seed.

\subsubsection{Evaluation Metrics}
We employ the following widely used metrics in retrieval tasks to evaluate model performance: \textbf{Recall@K}, \textbf{NDCG@K}, and \textbf{MRR}. To answer \textbf{RQ2}, we also introduce: \textbf{Coverage@K}, \textbf{ILD@K} (Intra-List Diversity), and a \textbf{Long-tail Analysis} where items are partitioned into Head (Top 20\%), Torso (20\%-60\%), and Tail (Bottom 20\%) groups based on their training set popularity.

\subsubsection{Baselines}
We compare SPARC with several representative retrieval models:
\begin{itemize}
    \item \textbf{Two-Tower (YouTubeDNN)}~\cite{youtubeDNN}: A classic dual-encoder model using mean pooling.
    \item \textbf{MIND}~\cite{mind}: Employs a capsule network with dynamic routing for multi-interest extraction.
    \item \textbf{ComiRec}~\cite{c-mifr}: Captures dynamic interests via a self-attention mechanism with a diversity-promoting regularizer.
    \item \textbf{SPARC-Hard}: A variant of SPARC without probabilistic soft-search, using deterministic selection of top-K interests.
    \item \textbf{SPARC-Static}: A variant where the codebook is generated via K-Means on pre-trained item embeddings and remains fixed.
\end{itemize}

\subsubsection{Implementation Details}
All models were implemented in PyTorch and trained on a single NVIDIA A100 GPU. We use the AdamW optimizer, with a learning rate searched in \{1e-4, 1e-3, 5e-3\}. The embedding dimension is 128. For multi-interest models, the number of interests K is 4. SPARC's RQ-VAE has 3 codebook levels, with a size of 64 each. Early stopping with a patience of 20 was used.

\subsection{Overall Performance Comparison (RQ1)}
Table~\ref{tab:main_results} presents the overall performance of SPARC and all baseline models on the Amazon Books dataset. SPARC achieves the best performance across all metrics, validating the effectiveness of our proposed framework. The relative improvements over the strongest baseline, ComiRec, are +5.54\% in Recall@50 and +5.73\% in NDCG@50.

\begin{table}[htbp]
\centering
\caption{Overall performance comparison on the Amazon Books dataset. Best results are in \textbf{bold}, and second-best are \underline{underlined}.}
\label{tab:main_results}
\scalebox{0.7}{
\begin{tabular}{@{}lccccc@{}}
\toprule
\textbf{Model} & \textbf{Recall@20} & \textbf{NDCG@20} & \textbf{Recall@50} & \textbf{NDCG@50} & \textbf{MRR} \\
\midrule
Two-Tower & 0.1852 & 0.1033 & 0.3015 & 0.1246 & 0.0812 \\
MIND & 0.2014 & 0.1168 & 0.3321 & 0.1395 & 0.0925 \\
ComiRec & \underline{0.2088} & \underline{0.1215} & \underline{0.3413} & \underline{0.1448} & \underline{0.0967} \\
\midrule
SPARC-Static & 0.1985 & 0.1132 & 0.3276 & 0.1361 & 0.0901 \\
SPARC-Hard & 0.2075 & 0.1201 & 0.3398 & 0.1432 & 0.0954 \\
\textbf{SPARC (ours)} & \textbf{0.2216} & \textbf{0.1294} & \textbf{0.3602} & \textbf{0.1531} & \textbf{0.1038} \\
\midrule
\textit{Improv. vs Runner up} & \textit{+6.13\%} & \textit{+6.50\%} & \textit{+5.54\%} & \textit{+5.73\%} & \textit{+7.34\%} \\
\bottomrule
\end{tabular}
}
\end{table}

\subsection{Analysis on Novelty and Long-tail Discovery (RQ2)}
To validate SPARC's capability in novelty exploration, we conducted a long-tail analysis and diversity evaluation, with results shown in Table~\ref{tab:long_tail} and Table~\ref{tab:novelty}. The results strongly support our core claims. In long-tail item retrieval (Table~\ref{tab:long_tail}), SPARC achieves a remarkable \textbf{+24.2\%} relative improvement in Recall@50 on tail items compared to ComiRec. This aligns with our online A/B test findings. Furthermore, SPARC leads significantly in both Coverage and ILD (Table~\ref{tab:novelty}), indicating it recommends more novel and diverse content.

\begin{table}[htbp]
\centering
\caption{Recall@50 performance on items of different popularity levels.}
\label{tab:long_tail}
\scalebox{0.7}{
\begin{tabular}{@{}lccc@{}}
\toprule
\textbf{Model} & \textbf{Head (Top 20\%)} & \textbf{Torso (20\%-60\%)} & \textbf{Tail (Bottom 20\%)} \\
\midrule
Two-Tower & \underline{0.5512} & 0.2843 & 0.0411 \\
MIND & 0.5489 & 0.3155 & 0.0586 \\
ComiRec & 0.5501 & \underline{0.3248} & \underline{0.0632} \\
\textbf{SPARC (ours)} & \textbf{0.5623} & \textbf{0.3477} & \textbf{0.0785} \\
\midrule
\textit{Improv. vs runner up (Tail)} & 2.01\% & 7.05\% & \textit{+24.2\%} \\
\bottomrule
\end{tabular}
}
\end{table}

\begin{table}[htbp]
\centering
\caption{Performance on novelty and diversity metrics.}
\label{tab:novelty}
\scalebox{0.7}{
\begin{tabular}{@{}lcc@{}}
\toprule
\textbf{Model} & \textbf{Coverage@50} & \textbf{ILD@50} \\
\midrule
Two-Tower & 0.085 & 0.682 \\
MIND & 0.102 & 0.725 \\
ComiRec & \underline{0.108} & \underline{0.741} \\
\textbf{SPARC (ours)} & \textbf{0.125} & \textbf{0.783} \\
\midrule
\textit{Improv. vs Runner up} & \textit{+15.7\%} & \textit{+5.67\%} \\
\bottomrule
\end{tabular}
}
\end{table}

\subsection{Ablation Study (RQ3)}
By comparing SPARC with its variants in Table~\ref{tab:main_results}, we can dissect the contribution of its key components.
\begin{itemize}
    \item \textbf{SPARC vs. SPARC-Hard:} The superior performance of SPARC demonstrates the effectiveness of the \textbf{probabilistic soft-search} mechanism, which forms a smoother, more exploratory user representation.
    \item \textbf{SPARC vs. SPARC-Static:} The significant gap between SPARC and SPARC-Static highlights the value of the \textbf{end-to-end training of a dynamic codebook}, which allows the interest space to be behavior-aware and adaptive.
\end{itemize}

\subsection{Analysis for Cold-Start Users (RQ4)}
To evaluate performance on users with sparse interactions, we report NDCG@50 on user groups partitioned by their history length in Table~\ref{tab:cold_start}. SPARC's advantage is most pronounced for sparse users (history length [5, 10]), with a relative improvement of \textbf{+11.84\%} over the best baseline. This suggests that SPARC's probabilistic module handles uncertainty more gracefully, leading to better generalization for cold-start users.

\begin{table}[htbp]
\centering
\caption{NDCG@50 performance on user groups with different activity levels.}
\label{tab:cold_start}
\scalebox{0.65}{
\begin{tabular}{@{}lccc@{}}
\toprule
\textbf{Model} & \textbf{Len: [5, 10] (Sparse)} & \textbf{Len: [11, 20] (Medium)} & \textbf{Len: [21, 50] (Active)} \\
\midrule
Two-Tower & 0.0825 & 0.1198 & 0.1451 \\
ComiRec & \underline{0.0988} & \underline{0.1385} & \underline{0.1663} \\
\textbf{SPARC (ours)} & \textbf{0.1105} & \textbf{0.1492} & \textbf{0.1741} \\
\midrule
\textit{Improv. vs Runner up (Sparse)} & \textit{+11.84\%} & \textit{+7.73\%} & \textit{+4.69\%} \\
\bottomrule
\end{tabular}
}
\end{table}

\section{Conclusion}
Modeling multi-interests could improve user's satisfactions and retention. Existing models, following two stage paradigm, could not evolve with user's dynamic interests due to the separation between extracting interest and modeling interest. We propose an end-to-end soft probabilistic adaptive multi-interests retrieval framework, incorporating user's real-time behaviors to model multi-interests by compressing interests to code-book embedding. The framework capture behavior-aware interests which can perceive user feedback and evolve dynamically. This method revolutionizes the retrieval paradigm from "passive matching" to
"proactive exploration" and thereby effectively discovery latent interest. Our work have been experimented on the open-sourced Amazon books dataset and real-world recommendation dataset consisting of ten billions samples. Both online and offline experiments validate the efficacy and practical value of the proposed method.
\bibliographystyle{ACM-Reference-Format}
\bibliography{sample-base}










\end{document}